\documentclass[12pt]{article}
\usepackage{graphicx,amsmath,amssymb}

\newlength{\dinwidth}
\newlength{\dinmargin}
\newcommand{\dif}{\mathrm{d}}
\newcommand{\diff}[1]{\frac{\mathrm{d}#1}{#1}}

\begin{document}
\titlepage
\begin{flushright}
  IPPP/09/78  \\
  DCPT/09/156 \\
  23rd November 2009 \\
\end{flushright}

\vspace*{0.5cm}

\begin{center}
  {\Large \bf NLO prescription for unintegrated parton distributions}

  \vspace*{1cm}
  \textsc{A.D. Martin$^a$, M.G. Ryskin$^{a,b}$ and G. Watt$^{a}$} \\

  \vspace*{0.5cm}                                                    
  $^a$ Institute for Particle Physics Phenomenology, University of Durham, DH1 3LE, UK \\
  $^b$ Petersburg Nuclear Physics Institute, Gatchina, St.~Petersburg, 188300, Russia
\end{center}                                                    
                                                    
\vspace*{1cm}                                                    
                                                    
\begin{abstract}
  We show how parton distributions unintegrated over the parton transverse momentum, $k_t$, may be generated, at NLO accuracy, from the known integrated (DGLAP-evolved) parton densities determined from global data analyses.  A few numerical examples are given, which demonstrate that sufficient accuracy is obtained by keeping only the LO splitting functions together with the NLO integrated parton densities.  However, it is important to keep the precise kinematics of the process, by taking the scale to be the virtuality rather than the transverse momentum, in order to be consistent with the calculation of the NLO splitting functions.
\end{abstract}

\section{Introduction}
Conventionally, hard processes involving incoming protons, such as deep-inelastic lepton--proton scattering, are described in terms of scale-dependent parton distribution functions (PDFs), $a(x,\mu^2) = xg(x,\mu^2)$ or $xq(x,\mu^2)$.  These distributions correspond to the density of partons in the proton with longitudinal momentum fraction $x$, integrated over the parton transverse momentum up to $k_t=\mu$.  They satisfy DGLAP evolution in the factorisation scale $\mu$, and are determined from global analyses of deep-inelastic and related hard-scattering data.  However, for semi-inclusive processes, parton distributions unintegrated over $k_t$ are more appropriate.  For example, unintegrated parton distributions play an important r\^ole in the description of the transverse momentum dependence of different inclusive hard processes, such as inclusive jet production in deep-inelastic scattering (DIS)~\cite{WMR}, electroweak boson production~\cite{WZ}, prompt photon production~\cite{pgam}, azimuthal correlations in high-$p_T$ dijet production~\cite{dijet}, etc.  Moreover, the exclusive cross sections for vector meson photoproduction~\cite{vm} or central exclusive diffractive Higgs boson production~\cite{cedh} are also calculated in terms of the unintegrated parton distributions.  In fact, so-called `$k_t$-factorisation' was originally established~\cite{ktfact} for heavy-quark pair production, so that the cross section for $pp\to Q\bar{Q}X$ is of the form:
\begin{equation} \label{eq:QQ}
  \sigma(pp\to Q{\bar Q}X) = \int\!\diff{x_1}\,\int\!\diff{x_2}\,\int\!\diff{k_{1,t}^2}\,\int\!\diff{k_{2,t}^2}\;f_g(x_1,k_{1,t}^2)\,f_g(x_2,k_{2,t}^2)\;\hat{\sigma}(\hat{s},M^2,k_{1,t}^2,k_{2,t}^2),
\end{equation}
where the $f_g$ are the gluon densities of the incoming protons, unintegrated over $k_{i,t}^2$, such that $f_g(x,k_t^2)(\dif{x}/x)(\dif{k_t^2}/k^2_t)$ is the number of gluons in the longitudinal and transverse momentum intervals from $x$ to $x+\dif{x}$ and from $k_t^2$ to $k_t^2+\dif{k_t^2}$, respectively, and $\hat{\sigma}$ is the $gg\to Q\bar{Q}$ subprocess cross section.

In general, the unintegrated distributions, $f_a(x,k_t^2,\mu^2)$, depend on two hard scales, $k_t$ and $\mu$, and so the evolution is much more complicated.  The additional scale $\mu$ plays a dual r\^ole.  On the one hand it acts as the factorisation scale, while on the other hand it controls the angular ordering of the partons emitted in the evolution.  In Refs.~\cite{WMR,KMR} a prescription was given which allows the unintegrated distributions to be determined from the well-known integrated distributions.  The prescription was based on the fact that due to strong $k_t$ ordering, inherent in DGLAP evolution, the transverse momentum of the final parton is obtained, to leading-order (LO) accuracy, just at the final step of the evolution.  Thus the $k_t$-dependent distribution can be calculated directly from the DGLAP equation keeping only the contribution which corresponds to a single real emission, while all the virtual contributions from a scale equal to $k_t$ up to the final scale $\mu$ of the hard subprocess are resummed into a Sudakov-like $T$-factor.  The factor $T$ describes the probability that during the evolution there are no parton emissions.

\begin{figure}
  \begin{center}
    \includegraphics[width=\textwidth]{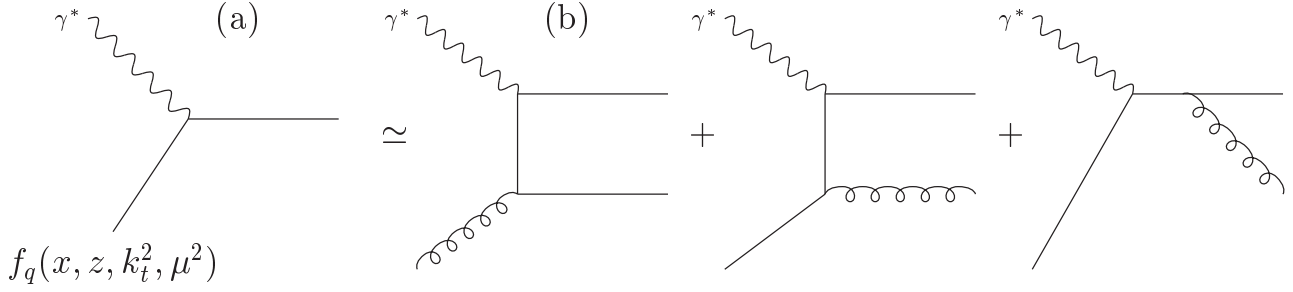}
    \caption{A schematic diagram of inclusive jet production in DIS at LO which shows the approximate equality between, on the left-hand side (a), the formalism based on the doubly-unintegrated quark distribution, $f_q(x,z,k^2_t,\mu^2)$, where the off-shell quark has virtuality $-k^2_t/(1-z)$, and on the right-hand side (b), the conventional QCD approach using integrated parton densities, $a(x,\mu^2)$, where the incoming partons are on-shell. \label{fig:LOjets}}
  \end{center}
\end{figure}
The idea is that, considering inclusive jet production in DIS, for example, the LO diagram at $\mathcal{O}(\alpha_{\rm em})$ computed using $k_t$-factorisation will already include, to a good approximation, the main effects (which are of kinematical origin) of the conventional LO QCD diagrams at $\mathcal{O}(\alpha_{\rm em}\,\alpha_S)$ computed using collinear factorisation.  This approximate equality is shown schematically in Fig.~\ref{fig:LOjets}.  The cross section for any hard process is then determined by convoluting the unintegrated parton distributions with the off-shell subprocess cross sections where the incoming partons have virtuality $-k_t^2$.  To be precise, it is necessary to also take account of the fraction $z$ of the light-cone momentum of the parent parton carried by the `unintegrated' parton, that is, to use `doubly-unintegrated' parton distributions~\cite{WMR}, $f_a(x,z,k^2_t,\mu^2)$, where the off-shell parton now has virtuality $-k^2_t/(1-z)$.  The doubly (or `fully') unintegrated distributions preserve the exact kinematics of the partonic subprocess (see also Ref.~\cite{collins}) and we speak of $(z,k_t)$-factorisation.  Here, we wish to extend the `last step' LO prescription for determining unintegrated parton distributions to next-to-leading order (NLO).\footnote{We do not consider \emph{fully} unintegrated distributions at NLO.  This would require the recalculation of the NLO DGLAP splitting kernels in fully unintegrated form, which is a necessary ingredient for a NLO parton shower Monte Carlo (see Ref.~\cite{jadach} for work in this direction).}

First, in Section~\ref{sec:lo} we recall the LO prescription, then we extend it to NLO in Section~\ref{sec:nlo}.  We show numerical results in Section~\ref{sec:num} and conclude in Section~\ref{sec:con}.  More details of the NLO derivation are given in an Appendix.

\section{LO prescription for unintegrated parton distributions} \label{sec:lo}
It is useful to review how LO unintegrated parton distributions, $f_a(x,k_t^2,\mu^2)$, may be calculated from the conventional (integrated) parton densities, $a(x,\mu^2)$, in the case of pure DGLAP evolution. As usual, we adopt a physical (axial) gauge, which sums over only the transverse gluon polarisations, so that the ladder-type diagrams dominate the evolution. Recall that the number of partons in the proton with longitudinal (or, to be precise, light-cone plus\footnote{The plus and minus light-cone components of a parton with 4-momentum $k$ are $k^\pm\equiv k^0\pm k^3$.}) momentum fraction between $x$ and $x+\dif x$ and transverse momentum $k_t$ between zero and the factorisation scale $\mu$ is $a(x,\mu^2)(\dif{x}/x)$, whereas the number of partons with longitudinal momentum fraction between $x$ and $x+\dif x$ and transverse momentum squared between $k_t^2$ and $k_t^2+\dif k_t^2$ is $f_a(x,k_t^2,\mu^2)(\dif{x}/x)(\dif{k_t^2}/k^2_t)$.  Thus the unintegrated distributions must satisfy the normalisation relation,\footnote{Note that the exact value of the upper limit in Eq.~\eqref{eq:norm}, and the possible non-logarithmic tail for $k_t>\mu$, are beyond NLO accuracy.}
\begin{equation} \label{eq:norm}
  a(x,\mu^2) = \int_0^{\mu^2}\!\diff{k_t^2}\,f_a(x,k_t^2,\mu^2),
\end{equation}
where $a(x,\mu^2) = xq(x,\mu^2)$ or $xg(x,\mu^2)$.  We start from the LO DGLAP equations evaluated at a scale\footnote{Usually DGLAP evolution is written in terms of the virtuality $k^2$, but at LO level this is the same.  The difference is a NLO effect.  We examine the difference in detail in Section~\ref{sec:num}.} $k_t^2$:
\begin{equation} \label{eq:DGLAPq}
  \frac{\partial\, xq(x,k_t^2)}{\partial \log k_t^2}  = \frac{\alpha_S(k_t^2)}{2\pi}\left[\,\sum_{b=q,g} \int_x^1\!\dif{z}\,P_{qb }(z)\,b  \left(\frac{x}{z},k_t^2 \right) - xq(x,k_t^2)\int_0^1\!\dif{\zeta}\;\,P_{qq}(\zeta ) \right ],
\end{equation}
\begin{equation} \label{eq:DGLAPg}
  \frac{\partial\, xg(x,k_t^2)}{\partial \log k_t^2}  = \frac{\alpha_S(k_t^2)}{2\pi}\left[\,\sum_{b=q,g} \int_x^1\!\dif{z}\,P_{gb }(z)\,b  \left(\frac{x}{z},k_t^2 \right) - xg(x,k_t^2)\int_0^1\!\dif{\zeta}\;\,(\zeta P_{gg}(\zeta )+ n_F P_{qg}(\zeta)) \right ],
\end{equation}
where $b(x,k_t^2) = xq(x,k_t^2)$ or $xg(x,k_t^2)$ and $P_{ab}(z)$ are the {\it unregulated} LO DGLAP splitting kernels.  The two terms on the right-hand sides of Eqs.~\eqref{eq:DGLAPq} and \eqref{eq:DGLAPg} correspond to real emission and virtual contributions respectively.  The virtual (loop) contributions may be resummed to all orders by the Sudakov form factor,
\begin{equation} \label{eq:Sudakovq}
  T_q (k_t^2,\mu^2) \equiv \exp \left (-\int_{k_t^2}^{\mu^2}\!\diff{\kappa_t^2}\,\frac{\alpha_S(\kappa_t^2)}{2\pi}\,\int_0^1\!\dif{\zeta}\; \,P_{qq}(\zeta ) \right ),
\end{equation}
\begin{equation} \label{eq:Sudakovg}
  T_g (k_t^2,\mu^2) \equiv \exp \left (-\int_{k_t^2}^{\mu^2}\!\diff{\kappa_t^2}\,\frac{\alpha_S(\kappa_t^2)}{2\pi}\,\int_0^1\!\dif{\zeta}\; \,(\zeta P_{gg}(\zeta )+n_F P_{qg}(\zeta)) \right ),
\end{equation}
which, recall, give the probability of evolving from a scale $k_t$ to a scale $\mu$ without parton emission.  Thus, from Eqs.~(\ref{eq:norm})--(\ref{eq:DGLAPg}), the unintegrated distributions have the form:
\begin{equation}
f_a(x,k_t^2,\mu^2) ~=~ T_a(k_t^2,\mu^2)\cdot\frac{\alpha_S(k_t^2)}{2\pi}\,\sum_{b=q,g}\,\int_x^1\!\dif{z}\,P_{ab }(z)\,b \left (\frac{x}{z}, k_t^2 \right).
\label{eq:fb}
\end{equation}
This equation can be rewritten as
\begin{equation}
f_a(x,k_t^2,\mu^2) ~=~ \frac{\partial}{\partial \log k_t^2}\left[\,a(x,k_t^2)\,T_a(k_t^2,\mu^2)\,\right],
 \label{eq:fa}
\end{equation}
since the derivative $\partial\,T_a/\partial\log k_t^2$ cancels the last terms in Eqs.~(\ref{eq:DGLAPq}) and (\ref{eq:DGLAPg}).

This definition is meaningful for $k_t > \mu_0$, where $\mu_0\sim 1$ GeV is the minimum scale for which DGLAP evolution of the conventional parton distributions, $a(x,\mu^2)$, is valid.  Integrating over transverse momentum up to the factorisation scale we find that
\begin{eqnarray}
  \int_{\mu_0^2}^{\mu^2}\!\diff{k_t^2}\,f_a(x,k_t^2,\mu^2) &=& \left[\,a(x,k_t^2)\,T_a(k_t^2,\mu^2)\,\right]_{k_t=\mu_0}^{k_t=\mu} \nonumber \\  &=& a(x,\mu^2) - a(x,\mu_0^2)\,T_a(\mu_0^2,\mu^2).
\end{eqnarray}
Thus, the normalisation condition Eq.~\eqref{eq:norm} will be exactly satisfied if we assume\footnote{A more complicated extrapolation of the contribution from $k_t<\mu_0$, which ensures continuity of $f_a$ at $k_t=\mu_0$, was given in Ref.~\cite{WZ}.}
\begin{equation} \label{eq:smallkt}
  \left.\frac{1}{k_t^2}\,f_a(x,k_t^2,\mu^2)\right\rvert_{k_t<\mu_0} = \frac{1}{\mu_0^2}\,a(x,\mu_0^2)\,T_a(\mu_0^2,\mu^2),
\end{equation}
so that the density of partons in the proton is constant for $k_t<\mu_0$ at fixed $x$ and $\mu$.

So far, we have ignored the singular behaviour of the unregularised splitting kernels, $P_{qq}(z)$ and $P_{gg}(z)$, at $z=1$, corresponding to soft gluon emission.  These soft singularities cancel between the real and virtual parts of the original DGLAP equations, Eqs.~\eqref{eq:DGLAPq} and \eqref{eq:DGLAPg}.  So it was enough to replace an explicit cutoff by the so-called `$+$' prescription.  Now, when real emission (which changes the $k_t$ of the parton) and the virtual loop contribution (which does not change the kinematics of the process) are treated separately, we must introduce the infrared cutoff explicitly.  The singularities indicate a physical effect that we have not yet accounted for.  It is the angular ordering caused by colour coherence~\cite{book}, which implies an infrared cutoff on the splitting fraction $z$ for those splitting kernels where a real gluon is emitted in the $s$-channel.  Indeed, the polar angle $\theta$ ordering 
\begin{equation}
  \ldots<\theta_{i-1}<\theta_i<\theta_{i+1}<\ldots
\end{equation}
implies the ordering of the ratios $\xi_i=p^-_i/p^+_i$, that is, of the rapidities $\eta_i=-(1/2)\log\xi_i=-\log\tan(\theta_i/2)$.  The light-cone components of the massless gluon-$i$ momentum ($p_i=k_{i-1}-k_i$) satisfy $p^-_i=p^2_{i,t}/p^+_i$, while the ratio of the momenta in the proton direction is given by the ratio of the momentum fractions carried by the $t$-channel gluons (see Fig.~2 of Ref.~\cite{WMR}), $z_i=k^+_i/k^+_{i-1}=x_i/x_{i-1}$.  If we denote $\bar p_i=p_{i,t}/(1-z_i)$ then we obtain
\begin{equation}
  \xi_i/\xi_{i-1}=(\bar p_i/z_{i-1}\bar p_{i-1})^2>1\;\;\;\mbox{that is}\;\;\;\; z_{i-1}\bar p_{i-1}<\bar p_i. 
\end{equation}
In the last step the angle is limited by the value of factorisation scale $\mu$.  If we choose $\mu=Q$ for DIS in the Breit frame, then the last inequality reads~\cite{WMR}
\begin{equation}
  z_n\bar p_n<\mu,
  \label{eq:angle}
\end{equation}
which leads to $z<\mu/(\mu+k_t)$ and provides the inequality $\theta_i<\theta_\mu$.

Thus the factorisation scale, $\mu$, is entirely determined from the kinematics of the subprocess at the `top' of the evolution ladder~\cite{WMR}.  So we define the infrared cutoff to be
\begin{equation}
  \Delta~\equiv~\frac{k_t}{\mu+k_t},
  \label{eq:D}
\end{equation}
then the precise expressions for the unintegrated quark and gluon distributions are
\begin{equation} \label{eq:a11}
  f_q(x,k_t^2,\mu^2) = T_q(k_t^2,\mu^2)\,\frac{\alpha_S(k_t^2)}{2\pi}\,\int_x^1\!\dif{z}\;\left[\,P_{qq}(z)\,\frac{x}{z}q\left(\frac{x}{z},k_t^2\right)\,\Theta(1-\Delta-z)+ P_{qg}(z)\,\frac{x}{z}g\left(\frac{x}{z},k_t^2\right)\,\right]
\end{equation}
and
\begin{equation}\label{eq:a12}
  f_g(x,k_t^2,\mu^2) = T_g(k_t^2,\mu^2)\,\frac{\alpha_S(k_t^2)}{2\pi}\,\int_x^1\!\dif{z}\;\left[\sum_q P_{gq}(z)\,\frac{x}{z}q\left(\frac{x}{z},k_t^2\right) + P_{gg}(z)\,\frac{x}{z}g\left(\frac{x}{z},k_t^2\right)\,\Theta(1-\Delta-z)\,\right].
\end{equation}

By unitarity the same form of the cutoff, $\Delta(\kappa_t)$, must be chosen in the virtual part. Thus we insert $\Theta(1-\Delta-\zeta)$ into the Sudakov factor for those splitting functions where a gluon is emitted in the $s$-channel and $\Theta(\zeta-\Delta)$ where a gluon is emitted in the $t$-channel.\footnote{The lower cutoff is beyond the LO accuracy and was simply introduced to make the formulation more symmetric.  It is not included in the NLO prescription of Section~\ref{sec:nlo}.}  Then
\begin{equation}
  T_q(k_t^2,\mu^2) = \exp\left(-\int_{k_t^2}^{\mu^2}\!\diff{\kappa_t^2}\,\frac{\alpha_S(\kappa_t^2)}{2\pi}\,\int_0^1\!\dif{\zeta }\,P_{qq}(\zeta )\Theta(1-\Delta-\zeta)\right).
  \label{eq:Tq}
\end{equation}
Recall that the exponent of the gluon Sudakov factor was already simplified by exploiting the symmetry $P_{qg}(1-\zeta)=P_{qg}(\zeta)$, so that the gluon Sudakov factor is
\begin{equation}
  T_g(k_t^2,\mu^2) = \exp\left(-\int_{k_t^2}^{\mu^2}\!\diff{\kappa_t^2}\,\frac{\alpha_S(\kappa_t^2)}{2\pi}\, \int^1_0\!\dif{\zeta }\;[~\zeta \,P_{gg}(\zeta )\Theta(1-\Delta-\zeta)\Theta(\zeta-\Delta) + n_F P_{qg}(\zeta)~]\right),
  \label{eq:Tg}
\end{equation}
where $n_F$ is the active number of quark--antiquark flavours into which the gluon may split.

Note that in the expression for the unintegrated distribution, Eq.~(\ref{eq:fa}), the derivative of $a$ with respect to $\log k^2_t$ gives the right-hand side of the usual DGLAP equation, which contains both the real emission (which we are looking for) and the virtual loop contribution. However the virtual contribution is cancelled by the derivative of $T_a$ with respect to $\log k^2_t$ .

In the next section we introduce an analogous prescription for unintegrated parton distributions which may be justified at NLO accuracy.

\section{Unintegrated parton distributions at NLO} \label{sec:nlo}
First, we discuss the structure of the NLO contribution. The original definition was that the amplitude was decomposed into the perturbative sum
\begin{equation}
  A~=~\sum^\infty_k \sum^\infty_{n=1}A_{n,k}\,\alpha_S^k\,(\alpha_S\log Q^2)^n,
\end{equation}
where $k=0,1,2,\ldots$ denote the LO, NLO, NNLO, $\ldots$ contributions.  However, nowadays the analogous decomposition is used for the anomalous dimensions, $\gamma_N$, which describe the evolution in terms of the OPE operators, e.g.~Mellin moments, $M_N$,
\begin{equation}
  M_N(Q)~=~M_N(Q_0)(Q^2/Q^2_0)^{\gamma_N},
\end{equation}
rather than for the amplitude.  That is, $M_N$ is written in the form
\begin{equation}
  M_N(Q)~=~M_N(Q_0)~\exp\left(\log(Q^2/Q^2_0)\sum_k c_k^N\alpha_S^k\right)~=~M_N(Q_0)~\sum_n\left(\log(Q^2/Q^2_0)\sum_k\frac{1}{n!} c_k^N\alpha_S^k\right)^n.
\end{equation}
\begin{figure}
  \begin{center}
    \includegraphics[height=8cm]{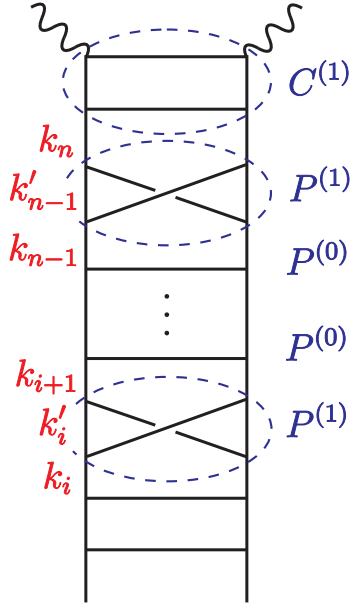}
    \caption[*]{The ladder diagram describing DGLAP evolution at NLO.  $C^{(1)}$ is the appropriate NLO coefficient function, and $P^{(0)}$ and $P^{(1)}$ are the appropriate LO and NLO splitting functions.}
    \label{fig:evol}
  \end{center}
\end{figure}
Thus the Feynman diagrams which describe DGLAP evolution have the structure shown in Fig.~\ref{fig:evol}. We have strong ordering of the transverse momenta, $k_{i,t} \gg k_{i-1,t}$, {\it between} the splitting functions, but no ordering in the transverse momenta of the $t$-channel partons which enter the NLO (NNLO, $\ldots$) splitting functions.

Here, we discuss NLO.  If we denote the transverse momentum in the NLO loop as $k'_{i,t}$, then we have the ordering
\begin{equation}
  k_{i,t} \ll k'_{i,t} \sim k_{i+1,t} \ll k_{i+2,t}.
\end{equation}
The `parton' of the $k_t$-factorisation approach at NLO is the $t$-channel quark or gluon placed between the coefficient and splitting functions. That is, the parton labelled as $k_n$ in Fig.~\ref{fig:evol}. There should be strong $k_t$ ordering between the loops which correspond to the NLO coefficient function and to the NLO splitting function. Otherwise the uppermost two loops should be assigned to the NNLO coefficient function.

We seek the prescription which gives the unintegrated $k_{n,t}$ distribution of the parton to NLO accuracy. To do this we have to extend the formalism of the last section in two steps. First, we have to replace the LO splitting functions $P_{ab}(z)$ by the corresponding `LO+NLO' splitting functions~\cite{CFP,FP} and the LO `global' parton densities $a$ by the NLO parton densities.  Secondly, we must take care of the precise value of the scale at which the parton densities $a$ are measured, and of the limits in the integrations over the terms which include the LO splitting functions. In the DGLAP evolution equation the current scale $k^2$ is not exactly equal to the parton transverse momentum $k_t$, instead we have\footnote{Eq.~(\ref{eq:slo}) assumes that the mass of the partonic system produced by the splitting is zero.  This is true at LO, but not at NLO, where a pair of massless partons may be created.  However, to account for non-zero mass of the pair, in the NLO splitting function, is a NNLO correction, which is beyond our NLO accuracy.}
\begin{equation}
  k^2=\frac{k^2_t}{1-z}.
  \label{eq:slo}
\end{equation}
At LO level, or at very small $z$, this scale difference is negligible.  However, to reach NLO accuracy we have to account for this effect, at least in the LO part of the splitting functions.  Thus, recalling Eq.~(\ref{eq:fb}), at NLO we have
\begin{equation}
  f^{\rm NLO}_a(x,k_t^2,\mu^2)~=~\int^1_x\!\mathrm{d}z\;T_a(k^2,\mu^2)\;
  \frac{\alpha_S(k^2)}{2\pi}\sum_{b=q,g}\tilde{P}_{ab}^{(0+1)}\left(z\right)\,b^{\rm NLO}\left(\frac{x}{z},k^2=\frac{k^2_t}{1-z}\right)\Theta (1-z-k^2_t/\mu^2),
  \label{eq:fNLO}
\end{equation}
where $\tilde{P}^{(0+1)}=\tilde{P}^{(0)}+(\alpha_S/2\pi) \tilde{P}^{(1)}$, and
\begin{equation}
  \tilde{P}^{(i)}_{ab}(z)~=~P^{(i)}_{ab}(z)-\Theta(z-(1-\Delta))~F_{ab}^{(i)}~\delta_{ab}~p_{ab}(z),
  \label{eq:Ptilde}
\end{equation}
with strength of the $z \to 1$ singularity in $P^{(i)}_{ab}$ given by~\cite{FP}
\begin{eqnarray}
  F_{qq}^{(0)}~=~C_F, & ~~~F_{qq}^{(1)}~=~-C_F\left(T_RN_F\frac{10}9+C_A(\frac{\pi^2}{6}-\frac{67}{18})\right), \\
  F_{gg}^{(0)}~=~2C_A, & ~~~F_{gg}^{(1)}~=~-2C_A\left(T_RN_F\frac{10}9+C_A(\frac{\pi^2}{6}-\frac{67}{18})\right).
\end{eqnarray}
Here $i=0,1$ denote the LO and NLO contributions, respectively, and $p_{qq}(z) = (1+z^2)/(1-z)$ and $p_{gg}(z)=z/(1-z)+(1-z)/z+z(1-z)$.  More details are given in the Appendix.

The last term in Eq.~(\ref{eq:Ptilde}) accounts for the coherence and eliminates the $1/(1-z)$ singularity in the splitting function caused by the emission of one soft gluon, which violates angular ordering Eqs.~(\ref{eq:angle},\ref{eq:D}). Recall that at NLO, the only strong singularity in $P_{ab}$ comes from the vertex and self-energy corrections to single gluon emission, see Table 1 of Ref.~\cite{CFP}. The contribution from the emission of two `real' gluons has (after the usual subtraction of that generated by two LO splitting functions) no $1/(1-z)$ singularity, but only the `soft' integrable term proportional to $\log(1-z)$.  Note that, in contrast to Eq.~(\ref{eq:fb}), at NLO $T_a$ and $\alpha_S$ occur inside the $z$ integration as $k^2=k^2_t/(1-z)$.

The final point is to consider the Sudakov factors\footnote{Note that, at NLO, the splitting function $P_{qq}$ contains terms which change the flavour of the quark, $P_{q_i q_j}$.  So now $T_q$ of Eqs.~(\ref{eq:Tq},\ref{eq:Tq1}) must include a sum over the flavour of the new quark $i$, and similarly for the sum over the initial quark $b=q_j$ in Eq.~\eqref{eq:fNLO}.} $T_a$ which resum the virtual DGLAP contributions during the evolution from $k^2$ to $\mu^2$.  At the NLO we get
\begin{equation}
  T_q(k^2,\mu^2) = \exp\left(-\int_{k^2}^{\mu^2}\!\diff{\kappa^2}\,\frac{\alpha_S(\kappa^2)}{2\pi}\,
    \int_0^1\!\dif{\zeta }\;~\zeta[\tilde{P}^{(0+1)}_{qq}(\zeta ) + \tilde{P}^{(0+1)}_{gq}(\zeta)]\right),
  \label{eq:Tq1}
\end{equation}
\begin{equation}
  T_g(k^2,\mu^2) = \exp\left(-\int_{k^2}^{\mu^2}\!\diff{\kappa^2}\,\frac{\alpha_S(\kappa^2)}{2\pi}\, \int^1_0\!\dif{\zeta }\;~\zeta [\tilde{P}^{(0+1)}_{gg}(\zeta ) + 2n_F\tilde{P}^{(0+1)}_{qg}(\zeta)]\right),
  \label{eq:Tg1}
\end{equation}
where, at NLO, we must take $\tilde{P}^{(0+1)}=\tilde{P}^{(0)}+(\alpha_S/2\pi) \tilde{P}^{(1)}$ and
\begin{equation}
  \Delta=\frac{\kappa_t}{\mu+\kappa_t}~~~~~~{\rm with}~~\kappa_t=\sqrt{\kappa^2(1-\zeta)}.
\end{equation}
The theta functions in Eq.~\eqref{eq:Ptilde} again provide the correct angular ordering for soft gluon emission.  If $\kappa_t \ll \mu$, the effect of $\Delta$ in the non-singular part of $\tilde{P}$ is negligible. The region of $\kappa_t$ close to $\mu$ should be specified by the convention of how to separate the NLO coefficient function (that is, the hard matrix element) and the unintegrated parton distributions of the $k_t$-factorisation approach. In particular, the diagram corresponding to the parton self-energy, or to large angle gluon emission, could be assigned either way; that is, to the parton distribution or to the coefficient function. Our $\Theta$ functions correspond to strong angular ordering. Our prescription means that the contribution coming from angles $\theta>\theta_\mu$ should be included in the coefficient function, while the lower $\theta$ domain should be assigned to the unintegrated parton distribution.  The angle $\theta_\mu$ is defined below Eq.~(\ref{eq:angle}).

\section{Numerical results at NLO} \label{sec:num}
For simplicity, we concentrate on the singlet evolution with $n_F=3$ and we use the corresponding MSTW 2008 PDFs~\cite{MSTW2008} and $\alpha_S$ evolved with three fixed flavours.
\begin{figure}
  \includegraphics[clip,width=\textwidth]{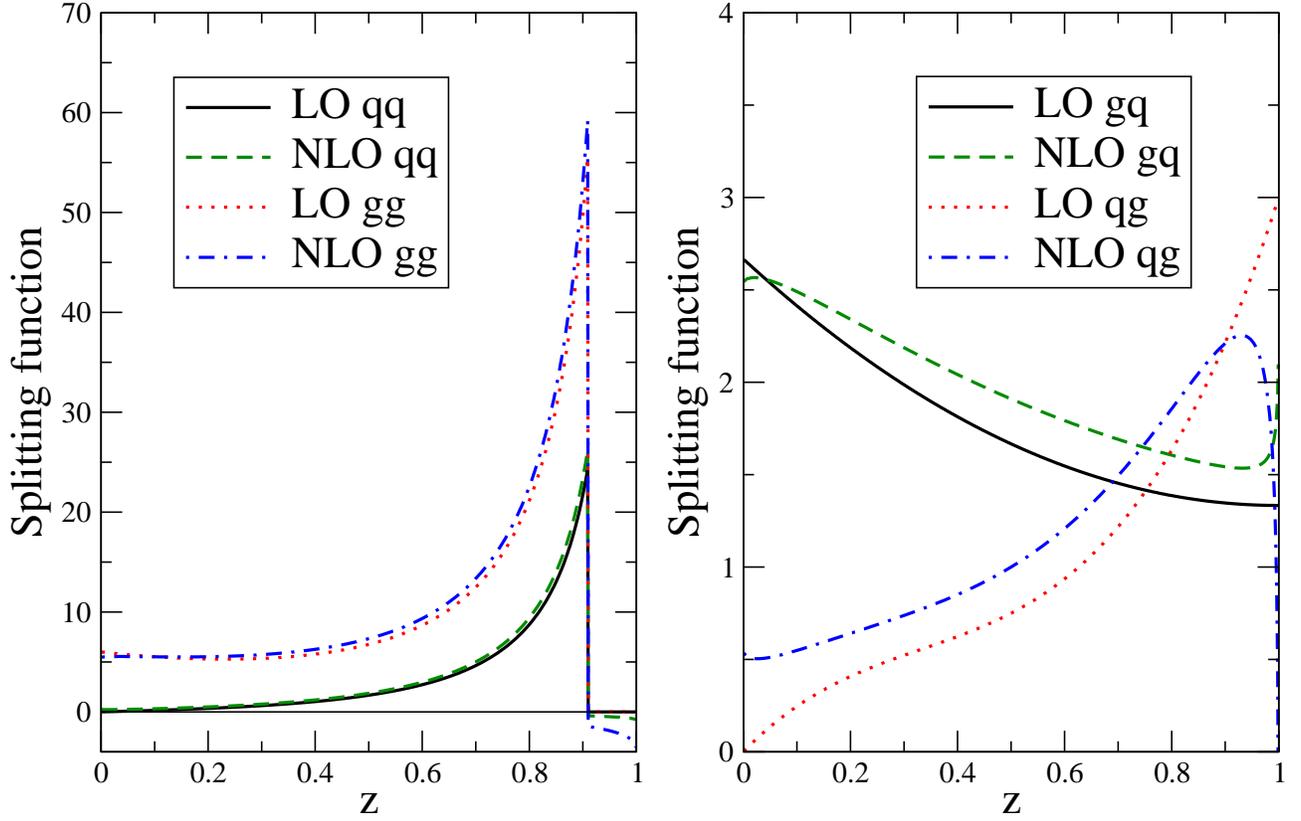} 
  \caption{DGLAP splitting functions, $z\tilde{P}_{ab}(z)$, given by Eq.~\eqref{eq:Ptilde}, at LO and NLO, after the subtraction for $z>\mu/(\mu+k_t)$ due to angular ordering, where we take $\mu^2 = 10^4$~GeV$^2$ and $k_t^2=100$~GeV$^2$.}
  \label{fig:split}
\end{figure}
First, in Fig.~\ref{fig:split} we show the DGLAP splitting functions at LO and NLO given by Eq.~\eqref{eq:Ptilde}.  It is seen that the NLO corrections to the splitting functions are relatively small in comparison with the LO contributions.  The largest NLO correction is perhaps in $P_{qg}$ at small $z$, where at LO there is no $1/z$ term as $z\to 0$.
\begin{figure}
  \includegraphics[clip,width=\textwidth]{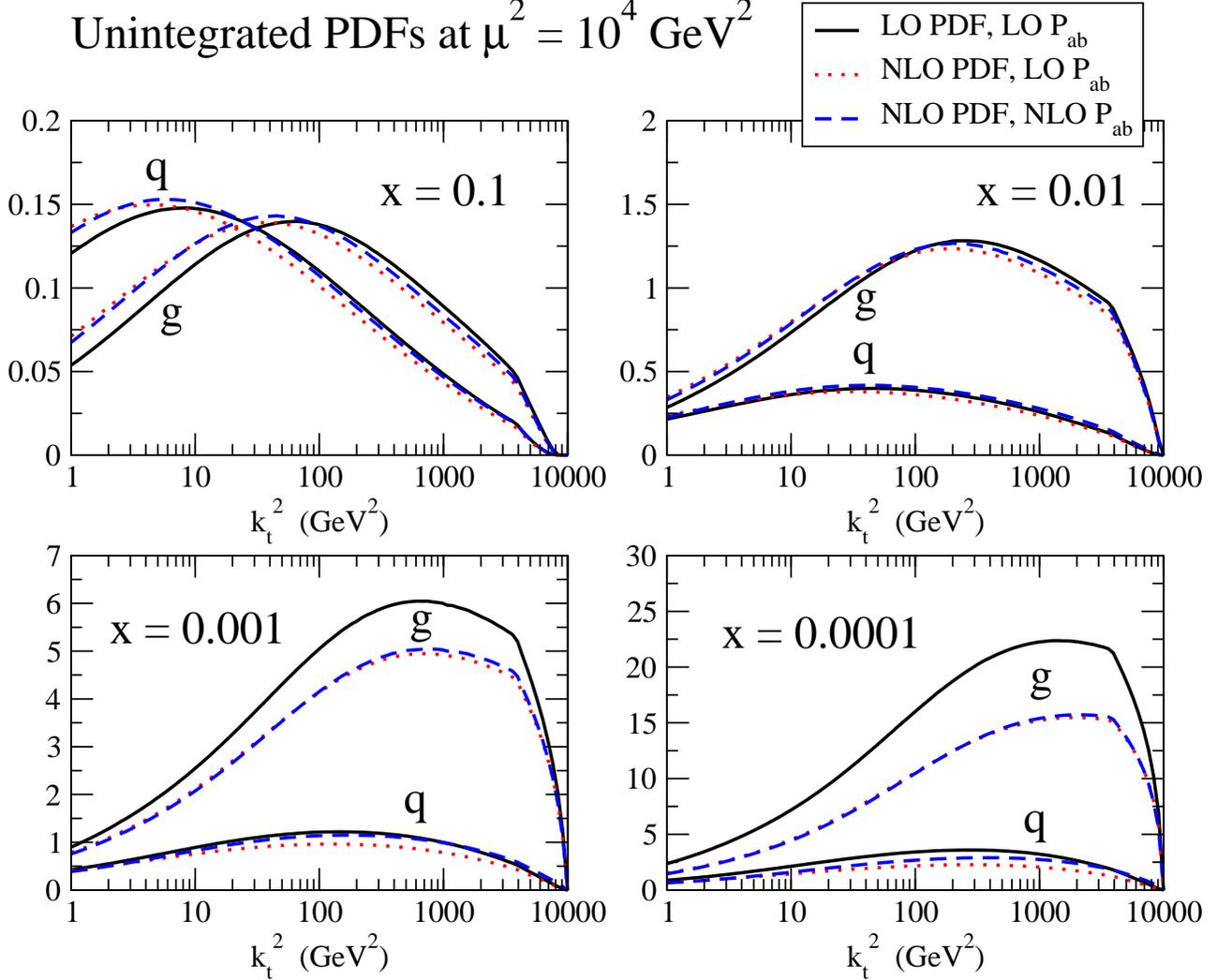}
  \caption{Unintegrated parton distributions, $f_a(x,k_t^2,\mu^2)$, given by Eq.~\eqref{eq:fNLO}, at $\mu^2 = 10^4$ GeV$^2$, for different orders of integrated PDF and `last-step' splitting function.  We use MSTW 2008~\cite{MSTW2008} integrated PDFs to generate the numerical predictions throughout this paper, except for Fig.~\ref{fig:f7} which shows the relative insensitivity to the choice of input PDF set.}
  \label{fig:f4}
\end{figure}
\begin{figure}
  \includegraphics[clip,width=\textwidth]{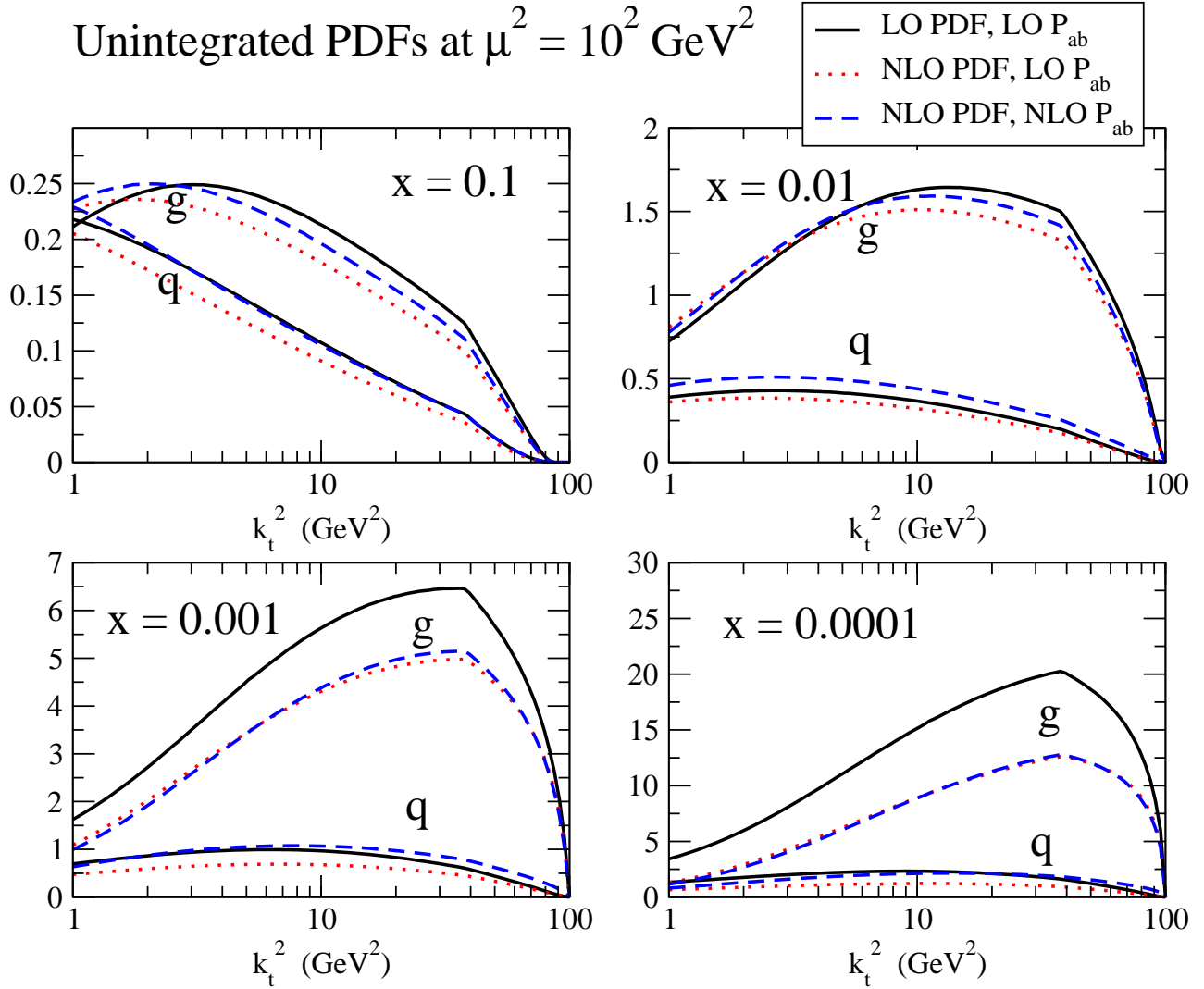}
  \caption{Unintegrated parton distributions, $f_a(x,k_t^2,\mu^2)$, given by Eq.~\eqref{eq:fNLO}, at $\mu^2 = 100$ GeV$^2$, for different orders of integrated PDF and `last-step' splitting function.}
  \label{fig:f5}
\end{figure}
In Fig.~\ref{fig:f4} we present the unintegrated PDFs as a function of $k_t^2$ at $\mu^2=10^4$~GeV$^2$ for $x=0.1,~0.01,~0.001$ and $0.0001$.  The corresponding plot at $\mu^2 = 100$~GeV$^2$ is shown in Fig.~\ref{fig:f5}.  Here, $q$ is the quark singlet distribution, $q=u+d+s+\bar{u}+\bar{d}+\bar{s}$.  Since the LO and NLO splitting functions are similar, the unintegrated PDFs calculated using the full NLO framework are close to the results keeping only the LO part of the splitting functions: compare the dotted and dashed curves in Figs.~\ref{fig:f4} and \ref{fig:f5}.  If we were to use the LO (rather than NLO) integrated PDFs, then we obtain the unintegrated distributions shown by the continuous curves.  We see that there is a sizeable enhancement of the unintegrated gluon at very small $x$, that is $x\lesssim 10^{-3}$.  This simply reflects the well-known difference between the integrated LO and NLO gluon distributions at small $x$.\footnote{The LO integrated gluon distributions obtained in the global analyses are larger in order to compensate for the absence of the $1/z$ pole in $P_{qg}$ at LO and the absence of a photon--gluon coefficient function at LO.}  The kink in the $k_t$ distributions at relatively large $k_t$ is due to the presence of two cutoffs in the expressions for the unintegrated distributions.  One cutoff, $\Theta(1-\Delta-z)$, accounts for the coherence effect, which leads to angular ordering.  The other cutoff is due to the bound on the virtuality, $k^2<\mu^2$, in the DGLAP evolution.  For $k_t \ll \mu$, the angular ordering of the soft gluon emissions is the stronger bound, while at large $k_t$ the bound on the virtuality takes over, with the transition point at $k_t=(1/2)(\sqrt{5}-1)\,\mu\simeq 0.6\mu$.

\begin{figure}
  \includegraphics[clip,width=\textwidth]{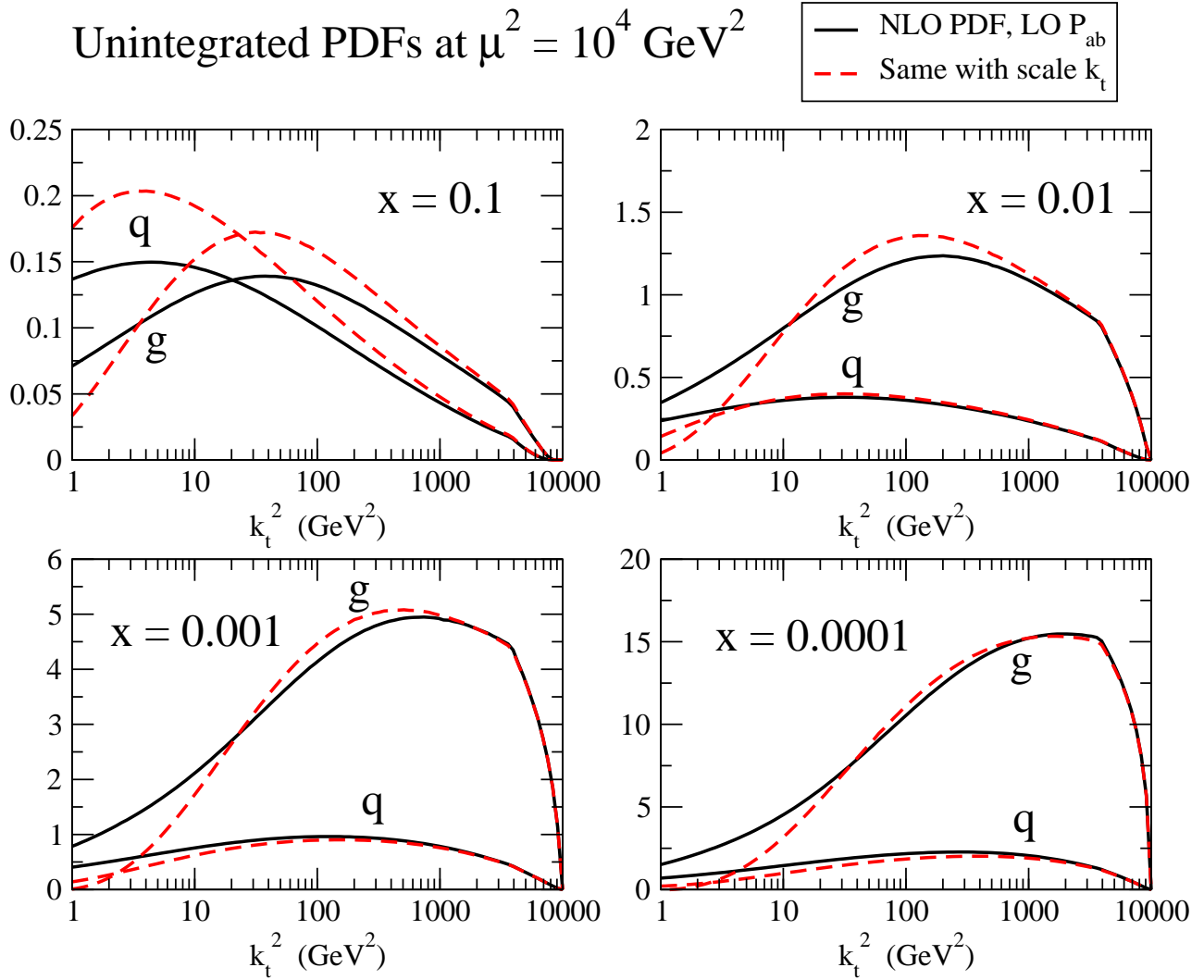}
  \caption{Unintegrated parton distributions, $f_a(x,k_t^2,\mu^2)$, given by Eq.~\eqref{eq:fNLO}, at $\mu^2 = 10^4$ GeV$^2$, compared to a prescription where the scale is taken as parton transverse momentum $k_t^2$ rather than virtuality $k^2=k_t^2/(1-z)$.  We use NLO PDFs with LO $P_{ab}$ in the last step in both cases.}
  \label{fig:f6}
\end{figure}
As noted above, the results obtained using the full NLO prescription (the dashed curves in Figs.~\ref{fig:f4} and~\ref{fig:f5}) are very close to that obtained simply using the LO splitting functions (dotted curves).  However, they differ from the results obtained from the old LO prescription~\cite{WMR,KMR}, described in Section 2, even when the NLO integrated PDFs are used in both cases.  The comparison is shown in Fig.~\ref{fig:f6}.  The difference arises because in the old LO prescription we did not care about the precise scale.  At LO accuracy, scales $k_t^2$ and $k^2$ are both acceptable.  However, at NLO accuracy, we must impose the correct scale, namely $k^2=k^2_t/(1-z)$, as in Eq.~(\ref{eq:slo}).  The main effect is the larger values of the Sudakov factors, $T$, due to larger lower limits in the integrals in Eqs.~(\ref{eq:Tq1},\ref{eq:Tg1}) leading to a smaller power in the exponent.  This enlarges the unintegrated PDFs for $k_t\ll \mu$.  For large $k_t$, however, the r\^ole of the $T$ factor is less important: $T$ is already close to 1.  On the other hand, a larger scale in the integrated PDFs and in $\alpha_S$ leads to somewhat smaller unintegrated densities, especially at relatively large $x$.  This is seen in Fig.~\ref{fig:f6}, where the dashed curve corresponds to the results obtained using the scale $k_t^2$, instead of $k^2$.

\begin{figure}
  \includegraphics[clip,width=\textwidth]{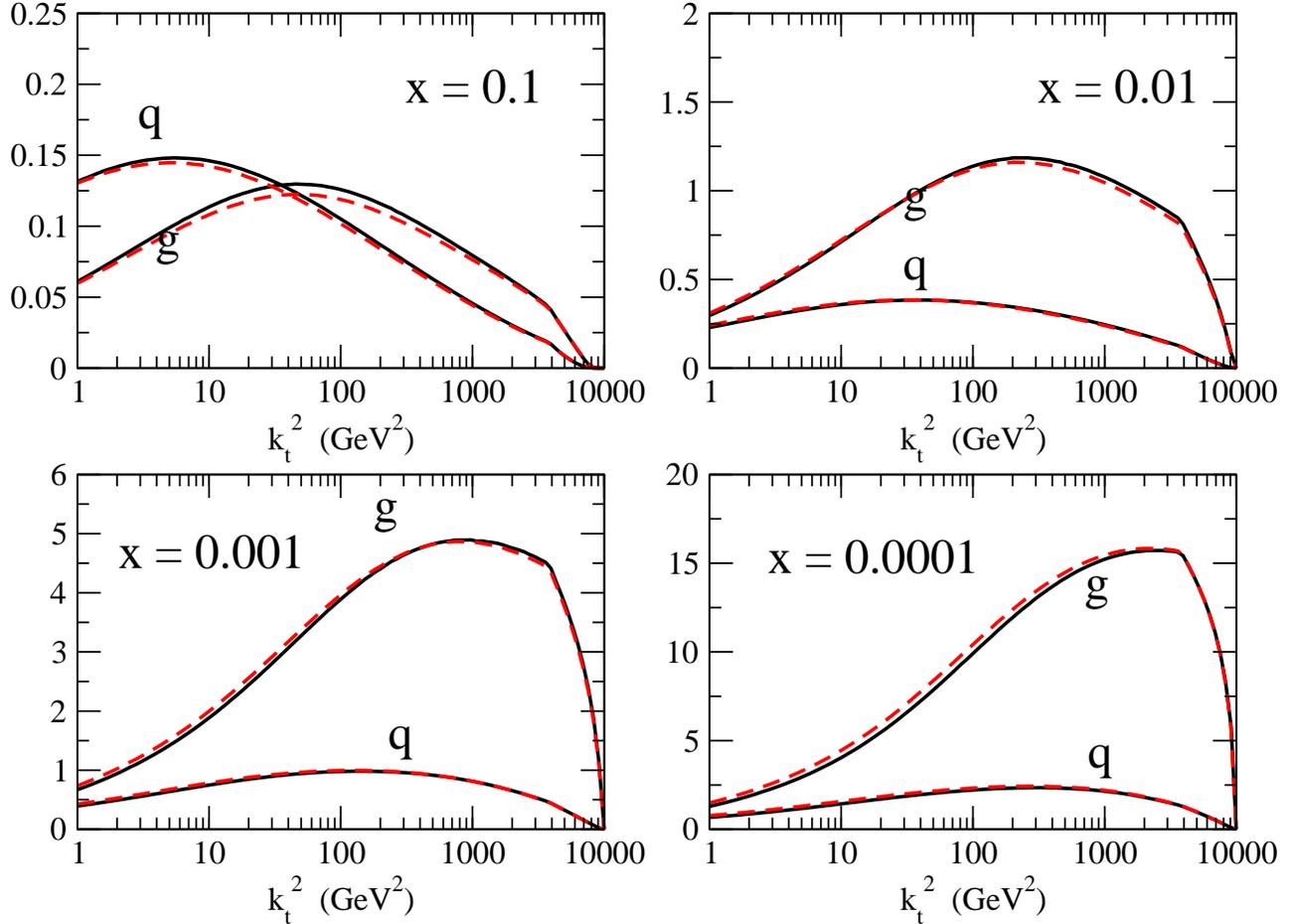}
  \caption{Unintegrated parton distributions, $f_a(x,k_t^2,\mu^2)$, given by Eq.~\eqref{eq:fNLO}, at $\mu^2 = 10^4$ GeV$^2$ for different input NLO PDFs: MSTW~\cite{MSTW2008} and CTEQ~\cite{CTEQ66} (both using LO $P_{ab}$ in the last step).  For comparison to CTEQ6.6, here we use the standard MSTW 2008 PDFs and $\alpha_S$ evolved with $n_F=5$, leading to a slight difference compared to the other plots in this paper where the PDFs and $\alpha_S$ are instead evolved with $n_F=3$.}
\label{fig:f7}
\end{figure}
We compare results obtained using MSTW 2008~\cite{MSTW2008} and CTEQ6.6~\cite{CTEQ66} PDFs in Fig.~\ref{fig:f7}, which shows that the predictions for NLO unintegrated PDFs are quite insensitive to different choices of the input integrated PDF set.

\begin{figure}
  \includegraphics[clip,width=\textwidth]{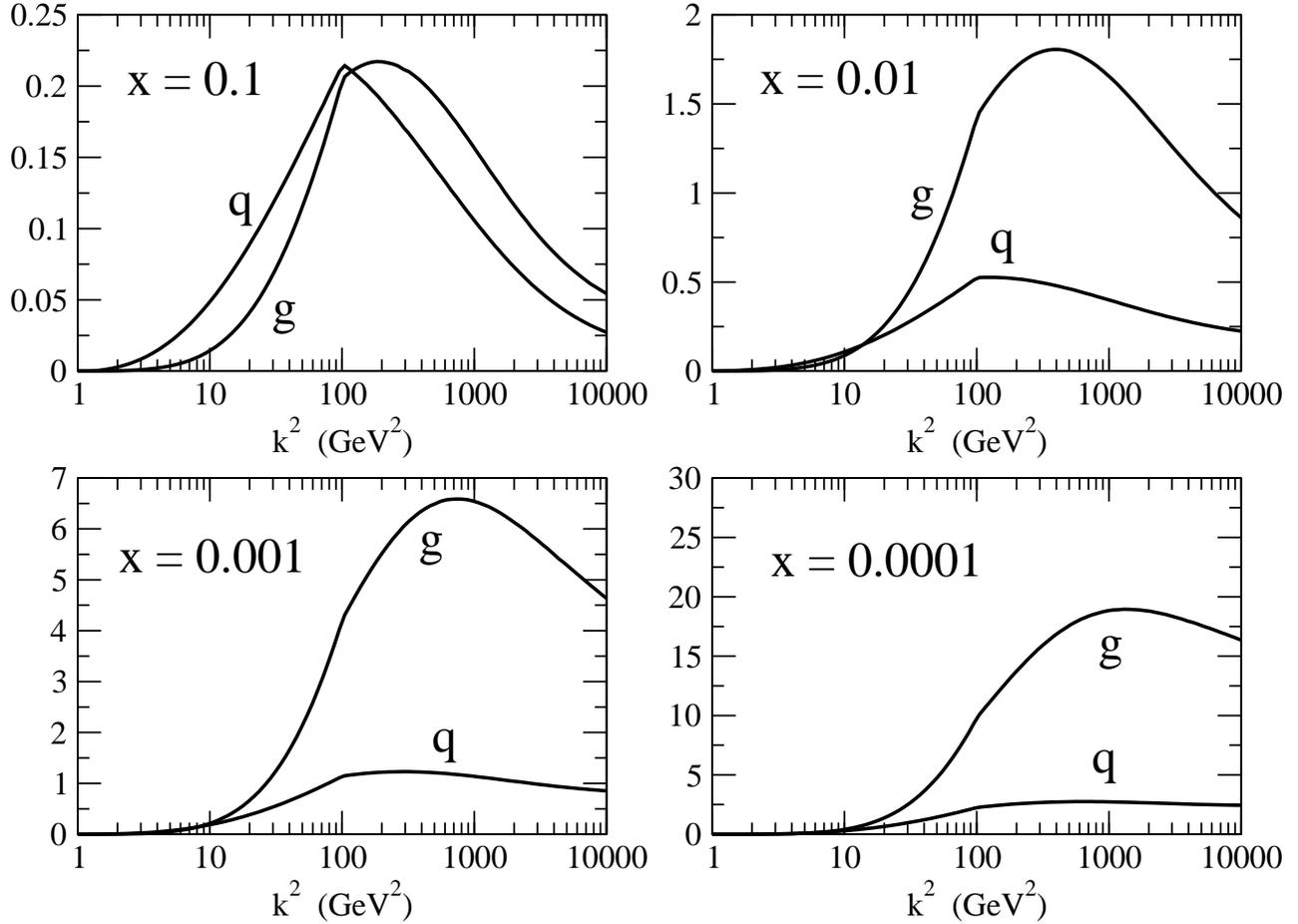}
  \caption{Unintegrated parton distributions, $f_a(x,k^2,\mu^2)$, given by Eq.~\eqref{eq:appvirt}, at $\mu^2 = 10^4$ GeV$^2$, defined as a function of the virtuality $k^2=k_t^2/(1-z)$.  We use NLO PDFs with LO $P_{ab}$ in the last step.}
\label{fig:f8}
\end{figure}
Usually the unintegrated distributions are defined as the density of partons at fixed $k_t$, and $k_t$-factorisation is used to calculate the observables.  In some specific cases, it may be advantageous to fix the virtuality $k^2$ rather than $k_t^2$ of the parton.\footnote{That is, in Eqs.~\eqref{eq:QQ} and \eqref{eq:norm}, $k_t^2$ is replaced by $k^2$, and we may speak of `$k$-factorisation' rather than $k_t$-factorisation.  Of course, in this case new coefficient functions would have to be defined and calculated corresponding to the unintegrated parton distributions at fixed $k^2$.  Note that `factorisation' is just a convenient way to write some amplitude or cross section in terms of the `Feynman diagram blocks' (in the physical planar gauge); see, for example, Fig.~\ref{fig:evol}.}  For completeness, we show an example of such unintegrated distributions in Fig.~\ref{fig:f8}.  Here we have used the LO splitting functions in the `last step' and NLO integrated PDFs, so the curves should be compared with the dotted lines in Fig.~\ref{fig:f4}.  Note that if we were to work in terms of $k^2$ then we would not have the last $\Theta$-function, $\Theta (1-z-k^2_t/\mu^2)$, which provides the inequality $k^2 <\mu^2$ in Eq.~(\ref{eq:fNLO}).  Instead we have to introduce an infrared cutoff $k_t>k_0=1$ GeV, that is $\Theta(k^2(1-z)-k_0^2)$.  (Physically, at large distances the system becomes colourless and so the $k_t$ is limited by the uncertainty principle.)  This infrared cutoff decreases the parton densities with fixed virtuality at low $k^2$, whereas the usual angular-ordering $\Theta$-function takes over for large $k^2$, again leading to a kink at the transition point where $k^2=k_0^2+k_0\,\mu$.

\section{Conclusions} \label{sec:con}

We have presented a prescription to generate unintegrated PDFs with NLO accuracy based on NLO integrated PDFs known from global parton analyses.  The prescription comes from a detailed consideration of the last step of the DGLAP evolution and accounts for the precise kinematics and the coherence effect (angular ordering) of the soft gluon emissions.  We compared the NLO and LO prescriptions and demonstrated that, to good accuracy, the NLO prescription can be simplified to keep only the LO splitting functions.  Nevertheless, the difference between the results obtained using the new NLO and the old LO prescriptions is not negligible (see Fig.~\ref{fig:f6}).  It turns out that it is important to account for the more precise kinematics in the LO splitting functions by taking the scale to be the virtuality $k^2$ rather than the transverse momentum $k_t^2$.  Since the virtuality $k^2>k^2_t$, the Sudakov suppression, which is driven at NLO by the value of $k^2$, is weaker now.  This leads to larger unintegrated PDFs in the region $k_t^2\ll\mu^2$.

Finally, a comment on the implications of these new predictions.  As a topical example we consider the use of the new unintegrated gluons in the prediction of the cross section of exclusive Higgs production, $pp \to p+H+p$, at the LHC.  The conclusion at the end of the last paragraph indicates that the exclusive Higgs cross section calculated to NLO accuracy should be larger than that calculated using the old LO prescription.  Note that the main contribution to the cross section~\cite{cedh} comes from the region $k^2_t \sim 4 ~{\rm GeV}^2$ $\ll M_H^2$ and $x \sim 0.01$, where the NLO prescription enhances the gluon density.  Since the predicted cross section is proportional to $f_g^4$, this enhancement has important implications for the viability of the exclusive Higgs signal at the LHC.  This is not the final effect, since diffractive Higgs production is driven by generalised skewed (not diagonal) unintegrated gluons.  The NLO skewed unintegrated PDFs can, in principle, be obtained by applying an analogous procedure to the NLO evolution of generalised PDFs.  At NLO we anticipate that the generalised PDFs will have a similar kinematical enhancement, but not so strong for the exclusive Higgs cross section, since the skewed distribution, $f_g(x,x',k^2_t,\mu^2)$, in the relevant region, $x^\prime\ll x$, is proportional to $\sqrt{T}$ and not to the full Sudakov factor $T$.

\appendix
\setcounter{equation}{0}
\renewcommand{\theequation}{A.\arabic{equation}}
\section*{Appendix}
Here we describe in more detail how the NLO prescription Eq.~(\ref{eq:fNLO}) follows from NLO DGLAP evolution.  The DGLAP splitting functions for singlet evolution up to NLO can be written as
\begin{equation} \label{eq:split}
  \mathcal{P}_{ab}^{(0+1)}\left(z\right) = \mathcal{P}_{ab}^{(0)}(z)+\frac{\alpha_S}{2\pi}\;\mathcal{P}_{ab}^{(1)}(z).
\end{equation}
The splitting function at each order ($i=0,1$) can be written as an unregularised part and a part proportional to $\delta(1-z)$, i.e.
\begin{equation}
  \mathcal{P}_{ab}^{(i)}(z) = P_{ab}^{(i)}(z) - \delta(1-z)\;K_a^{(i)}\;\delta_{ab},
\end{equation}
where the soft singularities as $z\to 1$ cancel between the two terms after convoluting with the parton distributions.  The coefficients of $\delta(1-z)$ can be obtained from conservation of momentum fraction, i.e.
\begin{equation}
  \sum_{b=q,g}\int_0^1\!\mathrm{d}\zeta\;\zeta\,\mathcal{P}_{ba}^{(i)}(\zeta) = 0 \qquad\Rightarrow\qquad K_a^{(i)} = \sum_{b=q,g}\int_0^1\!\mathrm{d}\zeta\;\zeta\,P_{ba}^{(i)}(\zeta).
\end{equation}
We define the Sudakov form factor as
\begin{equation} \label{eq:appsud}
  T_a(k^2,\mu^2) = \exp\left(-\int_{k^2}^{\mu^2}\!\frac{\mathrm{d}\kappa^2}{\kappa^2}\;\frac{\alpha_S(\kappa^2)}{2\pi}\;\sum_{b=q,g}\int_0^1\!\mathrm{d}\zeta\;\zeta\,P_{ba}^{(0+1)}\left(\zeta\right)\right),
\end{equation}
where
\begin{equation}
  P_{ab}^{(0+1)}\left(z\right) = P_{ab}^{(0)}(z)+\frac{\alpha_S}{2\pi}\;P_{ab}^{(1)}(z).
\end{equation}
The DGLAP equation for the integrated PDFs, $a(x,k^2)$, can be written as
\begin{align}
  \frac{\partial\,a(x,k^2)}{\partial\log k^2} &= \frac{\alpha_S(k^2)}{2\pi}\;\sum_{b=q,g}\int_x^1\!\mathrm{d}z\;\mathcal{P}_{ab}^{(0+1)}\left(z\right)\;b\left(\frac{x}{z},k^2\right)\\
&= \frac{\alpha_S(k^2)}{2\pi}\;\sum_{b=q,g}\left(\int_x^1\!\mathrm{d}z\;P_{ab}^{(0+1)}\left(z\right)\;b\left(\frac{x}{z},k^2\right)\;-\;a\left(x,k^2\right)\;\int_0^1\!\mathrm{d}\zeta\;\zeta\,P_{ba}^{(0+1)}\left(\zeta\right)\right)\\
&=\frac{\alpha_S(k^2)}{2\pi}\;\sum_{b=q,g}\int_x^1\!\mathrm{d}z\;P_{ab}^{(0+1)}\left(z\right)\;b\left(\frac{x}{z},k^2\right)\;-\;\frac{a\left(x,k^2\right)}{T_a(k^2,\mu^2)}\frac{\partial\,T_a(k^2,\mu^2)}{\partial\log k^2}.
\end{align}

The unintegrated parton distribution, as a function of the light-cone momentum fraction $x$, the virtuality $k^2$ and the factorisation scale $\mu^2$, is defined for $k^2<\mu^2$ as
\begin{align}
  f_a(x,k^2,\mu^2) &= \frac{\partial}{\partial\log k^2}\left[a(x,k^2)\;T_a(k^2,\mu^2)\right]\\
  &= T_a(k^2,\mu^2)\;\frac{\partial a(x,k^2)}{\partial\log k^2} + a(x,k^2)\;\frac{\partial T_a(k^2,\mu^2)}{\partial\log k^2}\\
  &= T_a(k^2,\mu^2)\;\frac{\alpha_S(k^2)}{2\pi}\;\sum_{b=q,g}\int_x^1\!\mathrm{d}z\;P_{ab}^{(0+1)}\left(z\right)\;b\left(\frac{x}{z},k^2\right). \label{eq:appvirt}
\end{align}
To get the unintegrated distribution as a function of transverse momentum we need to move the Sudakov factor and $\alpha_S$ inside the integral and account for the scale dependence Eq.~\eqref{eq:slo}, $k^2 = k_t^2/(1-z)$ using LO kinematics.  Then we obtain Eq.~\eqref{eq:fNLO}, i.e.
\begin{equation} \label{eq:appkt}
  f_a(x,k_t^2,\mu^2) = \int_x^1\!\mathrm{d}z\;T_a(k^2,\mu^2)\;\frac{\alpha_S(k^2)}{2\pi}\sum_{b=q,g}P_{ab}^{(0+1)}\left(z\right)\,b\left(\frac{x}{z},k^2\right)\;\Theta(\mu^2-k^2).
\end{equation}
Accounting for angular ordering, which additionally constrains the kinematics, in Eqs.~\eqref{eq:appsud}, \eqref{eq:appvirt} and \eqref{eq:appkt} we replace $P$ by $\tilde{P}$, as given in Eq.~\eqref{eq:Ptilde}.

\section*{Acknowledgements}
ADM and GW thank the UK Science and Technology Facilities Council for financial support.  MGR would like to thank the IPPP at the University of Durham for hospitality.  This work was supported by the grant RFBR 07-02-00023, by the Federal Program of the Russian State RSGSS-3628.2008.2.

\end{document}